\documentclass[final,12pt,letterpaper]{JHEP}
\usepackage{amsmath,amssymb}
\usepackage[final]{epsfig}

\allowdisplaybreaks[1]
\newcommand\skipthis[1]{{}}

\newcommand\ct[1]{{\sf {#1}},}
\newcommand\bt[1]{{\em {#1}},}
\newcommand\web[1]{{\tt {#1}}}
\newcommand\phepth[1]{{\tt [\hepth{#1}]}}

\chardef\til=`~

\newcommand\p{\ensuremath{\partial}}
\newcommand\evalat[2]{\ensuremath{\left.{#1}\right|_{#2}}}
\newcommand\abs[1]{\ensuremath{\left\lvert{#1}\right\rvert}}

\newcommand\field[1]{{\ensuremath{\mathbb{{#1}}}}}

\newcommand\vev[1]{{\ensuremath{\left\langle{#1}\right\rangle}}}

\newcommand\ep{\epsilon}
\newcommand\ov{\over}
\newcommand\V{{\cal V}}
\newcommand\vt{\vartheta}

\newenvironment{motivation}{\hfill \begin{minipage}{2in} \em}{
\end{minipage} \\}

\title{Stretched Strings in Noncommutative Field Theory}
\author{Hong Liu\thanks{\tt liu@physics.rutgers.edu} \ 
  and Jeremy Michelson\thanks{\tt jeremy@physics.rutgers.edu} \\ 
       New High Energy Theory Center \\ 
       Rutgers University \\
       126 Frelinghuysen Road \\
       Piscataway, NJ \ 08854}
\abstract{
Motivated by recent discussions of IR/UV mixing in noncommutative 
field theories, we perform a detailed analysis of 
the non-planar amplitudes of the bosonic open string in the presence of 
an external $B$-field at the one-loop level. We carefully isolate, at 
the string 
theory level, the contribution  which is responsible for the IR/UV behavior 
in the field theory limit. We  show that  it is a pure open string 
effect by deriving it  from the factorization of the one-loop amplitude into 
the disk amplitudes of intermediate {\it open\/} string insertions. 
We suggest that
it is natural to understand IR/UV mixing as the creation of intermediate 
``stretched strings''. 
}

\preprint{hep-th/0004013 \\ RUNHETC-00-13}

\begin{document}

\skipthis{
\begin{motivation}
``Laugh all you want.  I'm the one who's gonna be on T.V., lookin' all
buff''. \begin{center} --- Eric Cartman \end{center}
\end{motivation}
}

\section{Introduction}

Noncommutative field theory has recently received a lot of attention
from string theorists.
The appearance of noncommutative geometry in string theory was first
made clear in~\cite{cds,dh} (for earlier work see~\cite{miao}).
Subsequently, it was shown in~\cite{sw} that noncommutative gauge
theories will generically arise from open string theory in the presence of
a constant Neveu-Schwarz $B$-field.  More recently, it has been 
shown~\cite{mrs} that
perturbative noncommutative field theories~\cite{filk} 
exhibit an intriguing IR/UV mixing.

The mixing of IR and UV degrees of freedom
may be seen in many different ways. By thinking
of field theory quanta as pairs of dipoles moving in a magnetic 
field~\cite{sw,sj,bs,yin,mst}, one understands that
the relative position of the dipole is proportional 
to the center of mass momentum.  Alternatively, the mixing follows 
simply from the nature of the star product%
\footnote{$f(x)\ast g(x) \equiv \evalat{\exp\left[\frac{i}{2}
\Theta^{\mu\nu} \frac{\p}{\p x^\mu} \frac{\p}{\p {x'}^\nu}
\right] f(x)
g(x')}{x'=x}$, 
where $\Theta^{\mu\nu}$ is a constant antisymmetric matrix.}%
~\cite{mrs}; {\em e.g.}
\begin{equation}
\delta (x) \ast \delta(x) = \text{const}.
\end{equation}
The UV/IR mixing also allows for interesting soliton solutions that
would otherwise be prohibited by Derrick's theorem~\cite{gms}.

Perhaps one of the most interesting manifestations of the IR/UV mixing is the 
behaviour, first uncovered at one loop level 
in~\cite{mrs}, of non-planar loop integrals
in scalar field theories, which has since been generalized to 
higher loops~\cite{rs} and to many other theories~\cite{haya,grosse,
arefeva,abkr,texa,texas,mst,chu,bruce,rr}. 
One of the key elements in all these theories
is the emergence of a factor 
\begin{align} \label{thisfa}
&&
\exp[-{p \circ p \ov t}]; &\qquad p \circ p \equiv 
-p_{\mu} \left(\Theta^2\right)^{\mu \nu} p_{\nu},
\end{align}
in the integral over the Schwinger parameter $t$, 
after the loop momentum integration.
With non-zero external non-planar momentum $p$,
this factor regularizes otherwise divergent integrals. For this
reason, it has been suspected that
non-planar diagrams (not including planar sub-diagrams which can be 
renormalized) are finite; then the
whole theory
would be renormalizable. However, in~\cite{mrs} it was found that
there are actually
infrared singularities, associated with the factor~\eqref{thisfa} at $p=0$, 
in the diagrams that were UV divergent for
$\Theta=0$.
This is easily understood by thinking of $ p \circ p$ as a short distance
cutoff of the theory; taking $p=0$ corresponds to removing
the cutoff. If we introduce a UV cutoff $\Lambda$ in the theory, then 
the $p \rightarrow 0$ and $\Lambda \rightarrow \infty$ limits do not
commute and the theory does not appear to have a consistent 
Wilsonian description.   

It has been suggested in~\cite{mrs,rs} that the IR singularity may be 
associated with missing new light degrees of freedom in the theory. 
There it was shown that by adding 
new fields to the action with appropriate form,
the IR singularities in the 1-loop effective action of certain scalar 
field theories can indeed be reproduced and might result in a consistent
Wilsonian action. It was further suggested,  with the observation  that
$- \Theta^2$ looks like the closed string 
metric in the decoupling limit, that this procedure 
is  an analog of the open string theory factorization into closed string 
particles.

The purpose of this paper is two-fold. Since 
noncommutative field theory can generally be
reduced from string theory on a D-brane with a background 
$B$-field~\cite{dh,chekro,ardalan,chuho,scho,sw}
we try to identify the
string theory origin of the IR/UV behavior.  In particular, we 
examine whether
the above-mentioned effect can be attributed to closed string modes,
or  the
non-decoupling of other massive open string modes. Our other
motivation is that we 
would like to provide another intriguing example of IR/UV mixing 
at the string theory level. We point out that in the presence of a
$B$-field the
open string generically has  a finite winding mode proportional to its 
momentum. We argue that this effect persists at the field theory 
level,%
\footnote{This is rather similar to the dipole picture~\cite{sj,bs,yin}, 
and presumably
is important for the existence of T-duality in noncommutative
theories. See also~\cite{iikk,iso,ambjonf,ambjon} for discussions of 
non-local 
behaviors of noncommutative gauge theories via 
twisted large-$N$ matrix models and lattice formulations.} 
and  offers a simple explanation of the IR singularities of the non-planar 
diagrams. 

More explicitly, by examining the one-loop string amplitude and its 
reduction to field theory,
we will show that the occurrence of~\eqref{thisfa} and its associated 
IR singularities is an open string phenomenon,
and does not involve closed strings.%
\footnote{As this work was nearing completion, refs.~\cite{dorn,lee,bcr,shen}
appeared which also discussed one-loop string amplitudes and their 
reduction to noncommutative field theories. \skipthis{In particular 
\cite{lee} argues for a closed string interpretation, while
ref.~\cite{shen} argues for the open string interpretation.
We agree with~\cite{shen}.}} In particular we will see that the $- \Theta^2$
appearing in~\eqref{thisfa} does not originate from the closed string metric
at the string theory level.
Rather, if we write $x^\mu = \Theta^{\mu\nu} p_\nu$, then $p\circ p =
x^\mu G_{\mu\nu} x^{\nu}$, where $G_{\mu\nu}$ is the {\em open\/} string
metric.  We will argue in section~\ref{sec:iruv} that it is natural to 
interpret $x^\mu$ as a spacetime distance of a stretched string
and that therefore the
factor~\eqref{thisfa} is just $e^{-x^2/t}$, 
corresponding to a specific type of short-distance
regularization of the field theory.

This paper is organized as follows.  In section~\ref{sec:green}, we
introduce our notation, and write down the one-loop amplitude for
scattering of
open string tachyons in a background $B$-field.  To further illustrate
the structure of the amplitude, in
section~\ref{sec:ringfac}, we discuss the open string factorization
in the short cylinder limit.  In section~\ref{sec:disk}, we discuss
the closed string factorization in the long cylinder limit.
Having understood the structure of the amplitude, we are then able
to able to derive our understanding of the IR/UV mixing of~\cite{mrs},
in section~\ref{sec:iruv}.  
Section~\ref{sec:conclu} contains our discussion and conclusions.

\section{The One-Loop Amplitude} \label{sec:green}

In this section we  compute the one-loop scattering amplitudes for 
$M$ tachyons in the presence of a background $B$-field in the 
open bosonic string theory. 
We focus on the tachyons without Chan-Paton factors
for simplicity; however, since our arguments
and results can be traced to general properties of the Green function
and conformal field theory, we expect our conclusions to hold more generally.
We will mostly focus on the non-planar 
amplitudes, although the planar amplitudes can also be read from our results.
It is also easy to insert Chan-Paton factors.

The open string action is
(in conformal gauge)
\begin{equation} \label{action}
S = \frac{1}{4\pi\alpha'} \int d^2z \left\{g_{\mu\nu} \p_a X^\mu \p^a X^\nu
-i \varepsilon^{ab} B_{\mu\nu} \p_a X^\mu \p_b X^\nu \right\}
+ \frac{i}{2\pi\alpha'} \oint d\tau A_\mu \p_\tau X^\mu.
\end{equation}
We effectively set $F=dA$ to zero, by absorbing $F$ into $B$---this can be
considered either a notational convenience or the result of a gauge
transformation.
The action~\eqref{action} is exact in $\alpha'$, because we consider only
constant background fields.
This action leads to the boundary condition
\begin{equation} \label{bdyX}
\evalat{g_{\mu\nu}\p_n X^\nu - i B_{\mu\nu}\p_{\tau}X^\nu}{\text{bdy}} 
= 0,
\end{equation}
where $\p_n$ and $\p_\tau$ are respectively normal and tangential
derivatives.

The Green function for the bosonic open string on the annulus, in the
presence of a
$B$-field, was originally given in~\cite{at,acny,clny1}.  In the  
coordinate system in which the world-sheet is a flat annulus with
inner radius $a$ and unit outer radius (figure~\ref{annulus}),
it can be written as
\begin{equation} 
\label{green}
{\mathcal G}^{\mu\nu}(z,z') =
\vev{X^\mu(z) X^\nu(z')} =  -\frac{\alpha'}{2} 
\Bigl[ 
g^{\mu \nu} \ln \abs{Z}^2 + \Bigl(\frac{g-B}{g+B}\Bigr)^{\mu \nu}
\ln Y + \Bigl(\frac{g+B}{g-B}\Bigr)^{\mu \nu} \ln \Bar{Y}
\Bigr]
\end{equation}
where 
\FIGURE{
\includegraphics[height=1.5in]{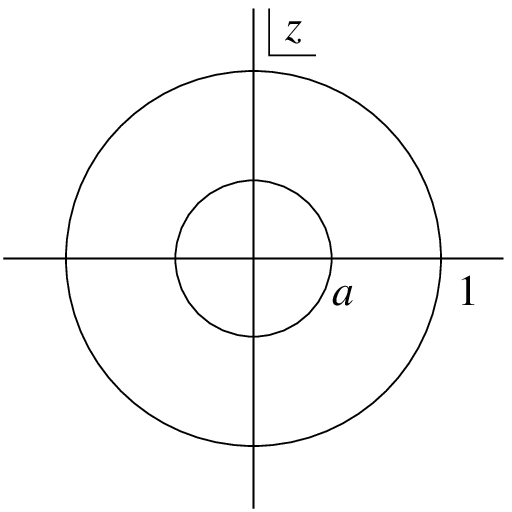}
\caption{The annulus. \label{annulus}}
}
\begin{subequations}
\begin{align}
Z  &= \frac{i a^{\frac{1}{6}}}{\eta(\tau)} \sqrt{z z'} 
\vartheta_{11}(\nu,\tau), &
Y  &= \frac{i a^{\frac{1}{6}}}{\eta(\tau)} \sqrt{z \Bar{z}'} 
\vartheta_{11}(\mu,\tau),
\end{align}
with
\begin{align}
a^2 &= e^{2\pi i\tau}, & e^{2 \pi i \nu} &= \frac{z'}{z}, &
e^{2 \pi i \mu} &= z \bar{z}'.
\end{align}
\end{subequations}
Note in~\eqref{green} the indices are raised and lowered by $g_{\mu \nu}$.
We closely follow the
conventions of~\cite{jp} for the definitions of the Jacobi-$\vartheta$
and Dedekind-$\eta$ functions.%
\skipthis{\footnote{Another notation is $\vartheta_1=\vartheta_{11}$
and $\vartheta_4=\vartheta_{01}$.}}

\begin{figure}[b]
\begin{center}
\includegraphics[height=1.5in]{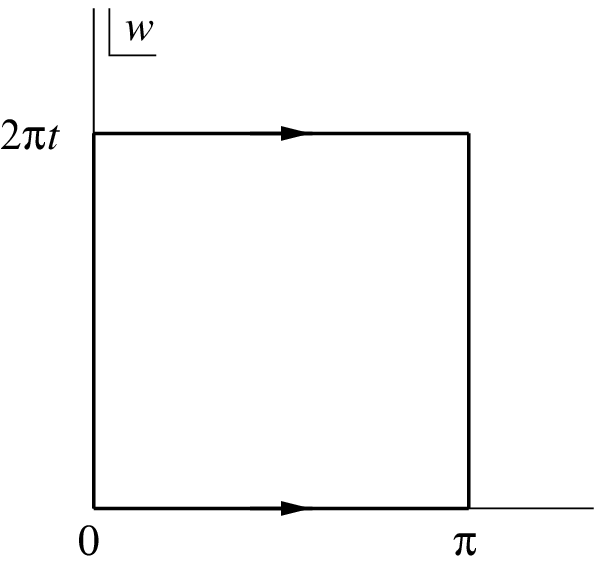}
\caption{The conformally equivalent description of the annulus as a
cylinder. \label{cylinder}}
\end{center}
\end{figure}

The Green function satisfies
\begin{subequations}
\begin{equation} \label{boxgreen}
\partial \Bar{\partial} {\mathcal G}_{\mu\nu}(z,z')
= -\pi \alpha' g_{\mu\nu} \delta^{(2)}(z-z'),
\end{equation}
and
\begin{equation} \label{bdygreen}
\left[
\partial_n {\mathcal G}_{\mu\nu}(z,z') 
- i B_\mu{^\lambda} \partial_\tau {\mathcal G}_{\lambda
\nu}(z,z')\right]_{\abs{z}=a,1} = \begin{cases} 0,&\abs{z}=a \\
-\alpha' g_{\mu\nu}, &\abs{z}=1 \end{cases}.
\end{equation}
\end{subequations}
\FIGURE{
\includegraphics[height=1.5in]{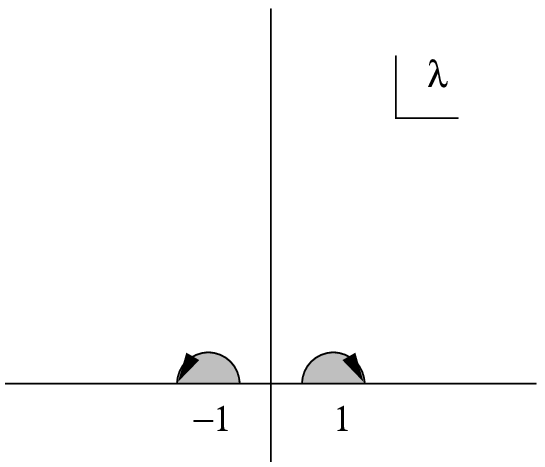}
\caption{The annulus as an upper half-plane, minus two regions with
identified boundaries. \label{half}}
}
\noindent
Equation~\eqref{bdygreen} is almost a consequence of
equation~\eqref{bdyX}; it differs from it on one boundary as a result
of Gauss' law, for otherwise equation~\eqref{boxgreen}
would be incompatible with equation~\eqref{bdygreen}.
We could follow~\cite{jp} and subtract a background charge from the
Green function instead of modifying the boundary condition; however,
this will not affect amplitudes.  Indeed, we will often in the
following, drop, with neither warning nor explanation, terms in the
Green function of the form
\hbox{\label{dropf+h} $f_{\mu\nu}(z,\Bar{z})+h_{\mu\nu}(z',\Bar{z}')$}, 
as momentum conservation
prevents such terms from contributing to on-shell amplitudes.

It will often be more convenient to use other coordinate systems.
The cylindrical coordinate (figure~\ref{cylinder}) is
\begin{equation} \label{defwcoords}
\begin{aligned}
w &= \sigma+i \tau
\end{aligned}; \qquad \begin{aligned}
\sigma &\in [0,\pi] \\
\tau &\in [0,2\pi t)
\end{aligned}; \qquad  \begin{aligned}
w \sim w+2\pi i t,
\end{aligned}
\end{equation}
where $\field{R}\ni t>0$ is the modulus. The relation between
the cylinder and the annulus is
\begin{align}
z&=e^{-{w \over t}}, &
a&= e^{-\frac{\pi}{t}}.
\end{align}

The coordinate on the upper half-plane is
\begin{equation} \label{deflambda}
\lambda = i \tan \frac{w-i \pi t}{2}.
\end{equation}
$\lambda$ maps the cylinder to the upper half plane with two
identified semi-circles removed, each being bounded by a segment of the real
axis---see figure~\ref{half}.
As $t\rightarrow\infty$, the semi-circles become infinitely small, and
centred at $\lambda=\pm1$.

When restricting to the boundary of the cylinder, the propagator
becomes (in cylindrical coordinates),
\begin{subequations}
\begin{align} 
\label{boungz}
& {\mathcal G}^{\mu\nu}(i\tau_1,i\tau_2) =
- \alpha' G^{\mu \nu} \ln \abs{
\vartheta_{11}\bigl(\frac{\tau_{12}}{2 \pi t},\frac{i}{t}\bigr)}^2
+ \frac{i}{2} \Theta^{\mu \nu} \bigl[\frac{\tau_{12}}{\pi t}
- \epsilon (\tau_{12} )\bigr], \\
\label{boung}
& {\mathcal G}^{\mu\nu}(\pi+i\tau_1,\pi+i\tau_2) 
= \alpha' g^{\mu \nu} {\pi \over t}
- \alpha' G^{\mu \nu} \ln \abs{
\vartheta_{11}\bigl(\frac{\tau_{12}}{2 \pi t},\frac{i}{t}\bigr)}^2
- \frac{i}{2} \Theta^{\mu \nu} \bigl[\frac{\tau_{12}}{\pi t}
- \ep(\tau_{12})\bigr], \\
\label{boungt}
& {\mathcal G}^{\mu\nu}(\pi+i\tau_1,i\tau_2) =
- \alpha' G^{\mu \nu} \ln \abs{
e^{- \frac{\pi}{4t}} 
\vartheta_{01}\bigl(\frac{\tau_1 - \tau_2}{2 \pi t},\frac{i}{t}\bigl)}^2,
\end{align}
\end{subequations}
where $\tau_{12} = \tau_1 - \tau_2$ and we have defined, following~\cite{sw}, 
\begin{subequations}
\begin{align} \label{defG}
G^{\mu \nu} & = \big ( {1 \ov g +B } g {1 \ov g -B} \big)^{\mu \nu} =
\big ( {1 \ov g +B }\big)_S^{\mu \nu}, \\ 
G_{\mu \nu} & = g_{\mu \nu} - (B g^{-1} B)_{\mu \nu}, \\
\label{deftheta}
\Theta^{\mu\nu} &= - (2 \pi \alpha')  
\big ( {1 \ov g +B } B  {1 \ov g -B} \big)^{\mu \nu} =
2 \pi \alpha' \big ( {1 \ov g +B } \big)_A^{\mu \nu}, 
\end{align}
\end{subequations}
where the subscript $S$ ($A$) denotes the (anti)symmetric part of the matrix.
Also, $\epsilon(\tau) = \tau/\abs{\tau}$ is the sign of $\tau$, and we
recall from {\em e.g.}~\cite{sw}
that
$g_{\mu\nu}$ is the closed string metric and
$G_{\mu\nu}$ is the open string metric (with inverse $G^{\mu\nu}$).
$\Theta^{\mu\nu}$ is an open string
noncommutativity parameter. Conversely, the closed string parameters
$g, B$ may be expressed in terms of the open string parameters $G, \Theta$
as
\begin{equation}
g + B =   \left( \frac{1}{G^{-1} + \frac{1}{2 \pi \alpha'} \Theta }
\right)_{\mu \nu}.
\end{equation}
In particular,
\begin{equation} 
\label{defg}
g^{\mu \nu} = G^{\mu \nu} - \left(\frac{1}{2 \pi \alpha'}\right)^2 
(\Theta G \Theta)^{\mu \nu}.
\end{equation}

We notice that the boundary Green functions on the cylinder have some
interesting
new features
compared to those of the
disk~\cite{acny,clny1} and those in the well-known case of $B=0$.
In~\eqref{boung}%
\footnote{The asymmetry of the Green function at the two boundaries is due to
our choice of asymmetric boundary conditions in~\eqref{bdygreen}, but
the net effect is independent of our choice.}
the dependence on the closed string metric does not drop out completely.  
We shall see later that this is essential for the non-planar 
amplitudes to factorize into the closed string channel  in the long
cylinder limit ($t \rightarrow 0$).
There is an additional dependence on $\Theta$ proportional to the
distance between
the two points. These linear 
terms will drop out in the planar amplitudes due to momentum
conservation  but will be important in non-planar amplitudes.
It is easy to see that planar amplitudes  will mimic those of
$B=0$ except for additional phase factors associated with external 
momenta.

Using the boundary Green functions~\eqref{boungz}--\eqref{boungt}, we now
write down the non-planar amplitude for $M$ tachyons with $N$ attached to
the  $\sigma=0$ boundary and  $M-N$ to the  $\sigma=\pi$ boundary. 
The vertex operators on the $\sigma=0$ boundary will be labeled by
$i, j, \dots=1, \dots, N$ and those on the 
$\sigma=\pi$ boundary by $r,s, \dots = N+1, \dots, M$.
$p,q$ go from $1$ to $M$. The normalized
open string tachyon vertex operator is
$\V_i = g_o e^{i k_{i\mu} X^{\mu}}$, where $g_o$ is the open 
string coupling.
We also define 
\begin{align}
k_i \cdot k_j &= k_{i \mu} G^{\mu \nu} k_{j \nu},&
k_i \times k_j &=  k_{i \mu} \Theta^{\mu \nu} k_{j \nu}.
\end{align}
The non-planar momentum---{\em i.e.}\ the momentum that flows between
the two boundaries---is
\begin{equation}
k = \sum_{i=1}^{N} k_i = - \sum_{r=N+1}^M k_r.
\end{equation}

In terms of the open string modular parameter $t$, the amplitude is
\begin{equation}
\begin{split} \label{ampop}
&S_{C^2} (1,\cdots, N;N+1, \cdots, M) 
= \int_{0}^{\infty}{dt \ov 2t} \vev{b(0) c(0)
\prod_{i=1}^{N} \oint \V_i(i \tau_i)   \prod_{r=N+1}^{M}
\oint \V_r(\pi + i \tau_r)}_{C^2} \\
& \quad = i \sqrt{\det G} \, g_o^M \, (2 \pi)^{d} \,  
\delta^{(d)} \Bigl(\text{\small $\sum_{p=1}^M$} k_p \Bigr)
\int_{0}^{\infty}\frac{dt}{2t} \, (8 \pi^2 \alpha' t)^{-\frac{d}{2}} 
\, \eta(it)^{-24} \, 
\left[\prod_{p=1}^{M} \int_{0}^{2 \pi t} d \tau_p \right]
\Psi_1 \Psi_2 \Psi_{12} \\*
& \quad \quad \times 
\exp\left[-\frac{\pi}{2t} \alpha' k_{\mu} (g^{-1}-G^{-1})^{\mu \nu}k_{\nu}
  \right]
e^{-\frac{i}{2}\sum_{i<j} (k_i \times k_j) [\frac{\tau_{ij}}{\pi t}
- \ep(\tau_{ij})]
+{i \over 2} \sum_{r<s} (k_r \times k_s) [\frac{\tau_{rs}}{\pi t}
- \ep(\tau_{rs})]},
\end{split}
\end{equation}
where we consider $d=26$, but equation~\eqref{ampop}
would hold on a
\hbox{D$(p{=}d{-}1)$-brane} with no transverse momentum, for general $d$, 
corresponding to a
$d$-dimensional field theory,
\begin{subequations}
\begin{gather}
\begin{align}
\Psi_1 &= \prod_{i<j} \abs{\psi_{ij}}^{2 \alpha' k_i \cdot k_j}, &
\Psi_2 &= \prod_{r<s} \abs{\psi_{rs}}^{2 \alpha' k_r \cdot k_s}, &
\Psi_{12} &= \prod_{i,r} (\psi^T_{ir})^{2 \alpha' k_r \cdot k_i},
\end{align} \\ \begin{align}
\psi_{ij}  &= - 2 \pi i \exp \left( -\frac{\tau_{ij}^2}{4 \pi t}\right)
\frac{\vt_{11}(i\frac{\tau_{ij}}{2 \pi},it)}{\vt_{11}'(0,it)}, &
\psi_{ir}^T &=  2 \pi \exp \left( -\frac{\tau_{ir}^2}{4 \pi t}\right) \,
\frac{\vt_{10}(i\frac{\tau_{ir}}{2 \pi}, it)}{\vt_{11}'(0, it)}.
\end{align}
\end{gather}
\end{subequations}

By a modular transformation of the $\vt$-functions  we can also
express the above amplitude in terms of the closed string modular 
parameter $s={\pi \ov t}$. This gives
\begin{equation} \label{ampcl}
\begin{split}
&S_{C^2} (1,\cdots, N;N+1, \cdots, M)
= i \sqrt{\det G} \, g_o^M \, (8 \pi^2 \alpha')^{-\frac{d}{2}} \, 
(2 \pi)^{d} \, \delta^{(d)} \Bigl(\text{\small $\sum_{p=1}^M$}k_p\Bigr) \\*
& \quad \times \int_{0}^{\infty}\frac{ds}{2} \,
\Bigl(\frac{\pi}{s}\Bigr)^{13-\frac{d}{2}}
\eta(\frac{is}{\pi})^{-24} \, \exp \big[-\frac{\alpha'}{2} s 
k_{\mu} g^{\mu \nu}  k_{\nu} \big]  
\left[\prod_{p=1}^{M} \int_{0}^{1} d \nu_p\right]
\Tilde{\Psi}_1 \Tilde{\Psi}_2 \Tilde{\Psi}_{12} \\*
& \quad \times 
\exp\left[
-\frac{i}{2} \sum_{i<j} (k_i \times k_j) [2 \nu_{ij} - \ep(\nu_{ij})]
+\frac{i}{2} \sum_{r<s} (k_r \times k_s) [2 \nu_{rs} - \ep(\nu_{rs})]
  \right],
\end{split}
\end{equation}
with
\begin{subequations}
\begin{gather}
\begin{align}
\Tilde{\Psi}_1 &= 
  \prod_{i<j} \abs{\tilde{\psi}_{ij}}^{2 \alpha' k_i \cdot k_j}, &
\Tilde{\Psi}_2 &= 
  \prod_{r<s} \abs{\tilde{\psi}_{rs}}^{2 \alpha' k_r \cdot k_s}, &
\Tilde{\Psi}_{12} &=  
  \prod_{i,r} (\tilde{\psi}^T_{ir})^{2 \alpha' k_r \cdot k_i},
\end{align} \\ \begin{align}
\tilde{\psi}_{ij}  &= 
\frac{\vt_{11}(\nu_{ij}, \frac{is}{\pi})}{\vt'_{11}(0, \frac{is}{\pi})}, &
\tilde{\psi}_{ir}^T  &= e^{-\frac{s}{4}}
\frac{\vt_{01}(\nu_{ir}, \frac{is}{\pi})}{\vt'_{11}(0, \frac{is}{\pi})},
\end{align}
\end{gather}
\end{subequations}
where for convenience we have defined $\nu_i = \frac{\tau_i}{2 \pi t}$.

We make several remarks:
\nopagebreak
\begin{enumerate}
\nopagebreak
\item \label{no-pi/t} \nopagebreak
$\psi$ and $\Tilde{\psi}$ have been defined to not
contain any factors of $\exp\bigl(\frac{\pi}{t}\bigr)$.
They do include $\vartheta$-functions, which encode the contributions
of intermediate
oscillation modes running in the loop,
and do not depend on the $B$-field.

\item \label{kixkj}
The  $k_i \times k_j$ contribution to the amplitude comes directly
from the $\Theta$ term in the propagators.

\item \label{diff-from-B0}
The structural differences of both~\eqref{ampop} and~\eqref{ampcl}
with those in the $B=0$ case indicate that the introduction of non-zero $B$
affects only the zero ({\em i.e.} non-oscillator) modes of the
intermediate state running in the loop.
This can also be deduced from the fact that in the mode expansion 
of $X^{\mu}$ on the disk, $B$ does not affect commutators involving the
oscillation modes.

\item \label{ftlimit}
In equation~\eqref{ampop} there is an exponential factor 
\begin{equation} \label{importf}
 \exp \left[-\frac{\pi}{2t} \alpha' k_{\mu} 
\bigl(g^{-1}-G^{-1}\bigr)^{\mu \nu}  k_{\nu}\right] 
=  \exp \left[\frac{1}{8 \pi \alpha' t} k_{\mu}(\Theta G \Theta)^{\mu \nu}
k_{\nu}\right],
\end{equation}
using equation~\eqref{defg}. 
The field theory limit of equation~\eqref{ampop} is obtained by taking
\begin{align} \label{field}
\alpha' &\rightarrow 0, &
t &\rightarrow \infty, &
\alpha' t &= \text{finite}, &
\alpha' \tau_p  &= \text{finite}.
\end{align}
In this limit the contributions from oscillators in~\eqref{ampop} 
drop out.  The open string modulus
$\alpha' t$ and vertex
coordinates $\alpha' \tau_p$ become the Schwinger parameters of the 
corresponding $\Phi^3$ field  theory with coupling constant
$\frac{g_o}{\alpha'}$.%
\footnote{See {\em e.g.}~\cite{di,mt}
for general discussions of calculating field theory amplitudes 
from string theory.}
The factor~\eqref{importf} does not arise in the $B=0$ case. But 
precisely this factor 
carries over to noncommutative field theory, where it regularizes 
otherwise divergent diagrams while introducing IR/UV mixing.
We emphasize that this is a very general phenomenon, not restricted to
the $\Phi^3$ amplitudes explicitly considered here.
For example, precisely the same factor appears if we replace the  
external legs  by vector vertex operators relevant to gauge theory
amplitudes, or if we reduce~\eqref{ampop} to $\Phi^4$ amplitudes using a
slightly different scaling~\cite{bcr}.  
The tensor with which the non-planar momenta are contracted,
\begin{equation} \label{pscm}
\tilde{g}^{\mu \nu}  = - (\Theta G \Theta)^{\mu \nu},
\end{equation}
was generally interpreted in noncommutative field 
theory discussions as the closed string metric in the 
$\alpha'\rightarrow0$ limit. 
However, we see here that
at the string theory level (finite $\alpha'$)
this is {\em not\/} the closed string metric, and it arises from {\em
open\/} string modes. In section~\ref{sec:iruv} we shall give an 
explanation of its origin.

\item \label{cl}
In terms of the closed string modulus, we have an exponential 
factor 
\begin{equation} \label{clofa}
 \exp \left[-\frac{\alpha'}{2} s 
\, k_{\mu} g^{\mu \nu}  k_{\nu}\right],
\end{equation}
in equation~\eqref{ampcl},
which indeed uses the closed string metric, in contradistinction
to the open string modulus expression~\eqref{importf}. 
This is essential for the factorization into disk amplitudes with a
closed string insertion, in the $s\rightarrow\infty$ limit.
\end{enumerate}

To have a precise understanding  of the structure of 
the amplitudes~\eqref{ampop} and~\eqref{ampcl} versus those  
in $B=0$ case, we now look at in detail  $t \rightarrow \infty$
and $s \rightarrow \infty$ limits. We shall see that 
the $\Theta$-dependent  parts of~\eqref{ampop} and~\eqref{ampcl} 
are completely determined by the factorization into open strings
and closed strings respectively.

\subsection{Open String Factorization} \label{sec:ringfac}

We shall expand equation~\eqref{ampop} about $t = \infty$, and verify that
it can be  rewritten in terms of a sum of disk amplitudes
with two (additional)
intermediate open string insertions. Historically, the
one-loop amplitudes were constructed in this way to ensure unitarity
(see {\em e.g.}~\cite{gsw}). 
Since, as we argued in remarks~\ref{no-pi/t} and~\ref{diff-from-B0},
$\Theta$-dependent
terms arise solely from the zero mode part of the intermediate operators,
it is enough to verify factorization to leading order, {\em i.e.} for
tachyon insertions.

We start by rewriting the amplitude~\eqref{ampop} in terms of 
coordinates on the upper half-plane [equation~\eqref{deflambda} and 
fig.~\ref{half}], for which the coordinate dependent part
of~\eqref{ampop} becomes
\begin{equation} \label{annuam}
\begin{split}
&(1-\lambda_1^2) 
\left(\prod_{i=2}^{N}  \int_{-\lambda_0}^{\lambda_0} d \lambda_i \right)
\left(\prod_{r=N+1}^{M} \int_{\frac{1}{\lambda_0}}^{-\frac{1}{\lambda_0}} 
d \lambda_r \right) \,  \prod_{p<q=1}^{M} \abs{(\lambda_p-\lambda_q) 
}^{2 \alpha' k_p \cdot k_q} \\
& \times \exp \Biggl[\frac{\alpha'}{2 \pi t} 
\Bigl(\text{\small $\sum_{p=1}^M$} k_p \tau_p\Bigr)^2 \Biggr]
\exp\Bigr[ 
\frac{i}{2} \text{\small $\sum_{r<s}$} (k_r \times k_s) 
    [\frac{\tau_{rs}}{\pi t} 
- \ep(\tau_{rs})]
-\frac{i}{2}\text{\small $\sum_{i<j}$}
   (k_i \times k_j) [\frac{\tau_{ij}}{\pi t}
- \ep(\tau_{ij})]\Bigl]
\end{split}
\end{equation}
where $\tau_p = - \ln\abs{\frac{\lambda_p +1}{\lambda_p -1}}+\pi t$ and 
$\lambda_0 = \tanh \frac{\pi t}{2}$. We have kept only the 
lowest order terms in the $t=\infty$ limit.
Note that in this form, aside from the $\Theta$-dependent terms, 
the amplitude is symmetric with respect to 
the external states on the two boundaries.

\begin{figure}[t]
\begin{center}
\includegraphics[width=5in]{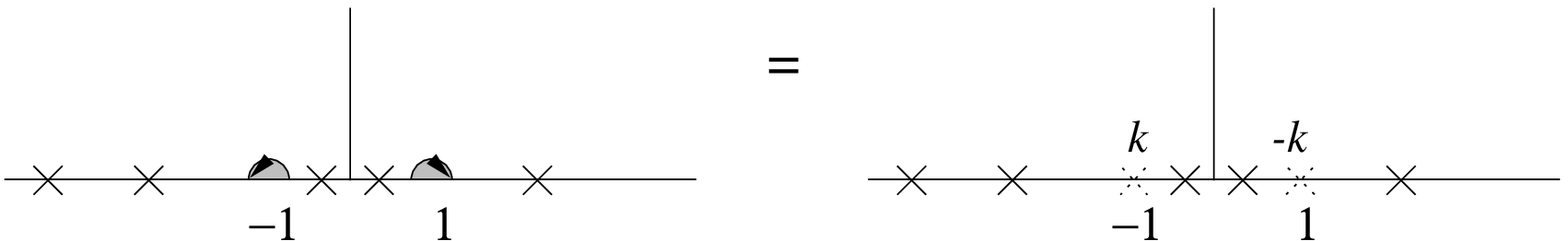}
\caption{As $t\rightarrow\infty$, the annulus diagram can be
approximated by the upper half-plane with momentum conserving tachyon
insertions at
$\lambda=\pm1$.  The leading-order disk amplitude is obtained by
contracting the two tachyon insertions only with each other;
contractions between external vertex operators and the insertions give
rise to the first-order corrections. \label{openfac}}
\end{center}
\end{figure}

The factorization formula can be obtained by inserting the full set 
of open string operators (including ghosts) at $\lambda=-1$ and $1$ of
the complex
$\lambda$-plane (see fig.~\ref{openfac}); to lowest order ({\em i.e.}
for a tachyon
insertion) it is given by~\cite{jpmod,jp}
($q = e^{-2 \pi t}$)
\begin{equation} \label{diskam} \raisetag{3\baselineskip}
\begin{split}
S & =  {1 \ov  K_o}\,  
g_o^M \int_{0}^{1} \frac{d q}{q} \int \frac{d^dk}{(2 \pi)^d}
\, q^{\alpha'(k^2 + m^2)} \left[\prod_{p=2}^M \int d \lambda_p \right] \\
& \times \vev{c^1 e^{ik \cdot X}(-1)
c^1 e^{ik_1 \cdot X} (\lambda_1) \,\,
\left[ \prod_{i=2}^{N} e^{ik_i \cdot X} (\lambda_i) \right]
b_0 c^1  e^{-ik \cdot X}(1)  
\left[\prod_{r=N+1}^{M} e^{ik_r \cdot X} (\lambda_r) \right] }_{D_2}
\end{split}
\end{equation}
where $K_o =1/(g_o^2 \alpha')$ normalizes the tachyon 
two-point  function on the disk, and
the measure of $q$ integration comes from 
the sewing of the two half circles in figure~\ref{openfac}.

Recall that the disk Green function on the upper half-plane is given
by~\cite{sw}
\begin{equation} \label{Gdisk}
{\mathcal G}^{\mu \nu} (\lambda_i,\lambda_j) 
= - \alpha' G^{\mu \nu} \ln \abs{\lambda_i
-\lambda_j}^2 + \frac{i}{2} \Theta^{\mu \nu} \epsilon(\lambda_i-\lambda_j);
\end{equation}
then the result of~\eqref{diskam} can be easily 
written down. To keep our formula from becoming too long we will
omit the
modulus and coordinate integrations and unimportant prefactors, and write
\begin{equation} \label{factf}
S  = 
(1- \lambda_1^2) \left[\prod_{p<q=1}^{M} 
\abs{\lambda_p-\lambda_q}^{2 \alpha' k_p \cdot k_q}\right]  \, 
e^{\frac{i}{2}\sum_{i<j=1}^{N} (k_i \times k_j)\ep(\tau_{ij}) -  
{i \over 2} \sum_{r<s=N+1}^{M} (k_r \times k_s) \ep(\tau_{rs})} 
\times I
\end{equation}
with 
\begin{equation}
\begin{split} \label{intek}
I & = \int \frac{d^dk}{(2 \pi)^d}
e^{- 2 \pi t \alpha'k^2} \exp \Bigl[-\text{\small $\sum_{p=1}^{M}$} 2 \alpha' 
k \cdot (k_p \tau_p) 
   + i \text{\small $\sum_{i=1}^{N}$} k_{\mu} \Theta^{\mu \nu}
k_{i \nu} \Bigr] \\
& = \sqrt{\det G} (8 \pi^2 \alpha' t)^{-\frac{d}{2}}
 \exp \left[\frac{1}{8 \pi \alpha' t} \Bigl(\sum_{i=1}^N k_{i \mu}\Bigr)
(\Theta G \Theta)^{\mu \nu} \Bigl(\sum_{j=1}^N k_{j \nu}\Bigr)\right] \\*
& \quad \times  \exp \left[\frac{\alpha'}{2 \pi t} 
  \Bigl(\sum_{p=1}^M k_p \tau_p\Bigr)^2 \right]
e^{-\frac{i}{2 \pi t}\sum_{i<j=1}^N (k_i \times k_j) \tau_{ij}
+ \frac{i}{2 \pi t} \sum_{r<s=N+1}^M (k_r \times k_s) \tau_{rs}}
\end{split}
\end{equation}
It is can be checked  that by combining equations~\eqref{intek} 
and~\eqref{factf}
and restoring the prefactors in~\eqref{factf} we precisely
recover~\eqref{annuam}.
Since the oscillator parts of higher order operators
always involve derivatives of the $X$, 
their contraction with
external states will not result in any additional dependence on
$\Theta$\skipthis{, and
there is no new term in the $k$ integral other than those already 
in~\eqref{intek}}.
So the contributions from oscillator modes follow exactly as in the 
$B=0$ case, while the new factors,~\eqref{importf} and the last exponential
in~\eqref{intek}, come from the open string zero modes.

\subsection{Closed String Factorization} \label{sec:disk}

We examine the amplitude~\eqref{ampcl} in the closed string limit
$t\rightarrow0$, or $s\rightarrow\infty$. It is easy to see
that~\eqref{ampcl} factorizes to closed string insertions on the disk in
precisely the same manner as in the $B=0$ case.  This follows since
the $\Theta$-dependent factors are already factorized with respect
to the two boundaries.
Thanks to the factor~\eqref{clofa}, the 
closed string poles are indeed located according to the mass-shell
condition in terms of the closed string metric.
Here we shall demonstrate that the $\Theta$-dependent factors in
the third line of~\eqref{ampcl} arise from the zero mode part of the
closed string insertion. For this purpose it is again enough to look at
the (closed string) tachyon insertion.
The disk Green function on the unit circle in figure~\ref{annulus} is given
by,
\begin{equation} \label{disgr}
{\cal G}^{\mu \nu}(\theta_i, \theta_j) = - \alpha' G^{\mu \nu} 
\ln \abs{2 \sin\Bigl(\frac{\theta_{ij}}{2}\Bigr)}^2 
+ \frac{i}{2} \Theta^{\mu \nu}
\Bigl[\frac{\theta_{ij}}{\pi} - \epsilon(\theta_{ij})\Bigr].
\end{equation}
For pure open string amplitudes, the terms linear in
$\theta_{ij}\equiv\theta_i-\theta_j$
in~\eqref{disgr} cancel by momentum conservation. However when there 
is a closed string insertion the momentum conservation no longer causes
it to vanish, but gives a 
contribution precisely as in the third line of~\eqref{ampcl}.

Closed string factorization also gives an alternative derivation of the 
relation between closed and open string couplings. 
The relation was first derived
in~\cite{sw} by arguing the equivalence of the Born-Infeld actions
expressed in terms of closed and open string parameters. Since numerical 
factors follow precisely as in $B=0$ case, here we shall only focus
on the metric dependence. For tachyon insertion
the factorization takes the form, (for convenience we take $d=26$)
\begin{equation} \label{ampcld2d2}
S_{\text{annulus}} = \frac{1}{K_{c}}\,
S_{D_2}(1,\dots,N,k) \, \frac{1}{k_\mu g^{\mu\nu} k_\nu - 4} \,
S_{D_2}(N+1,\dots,M,-k),
\end{equation}
\begin{figure}[t]
\begin{center}
\includegraphics[width=4.0in]{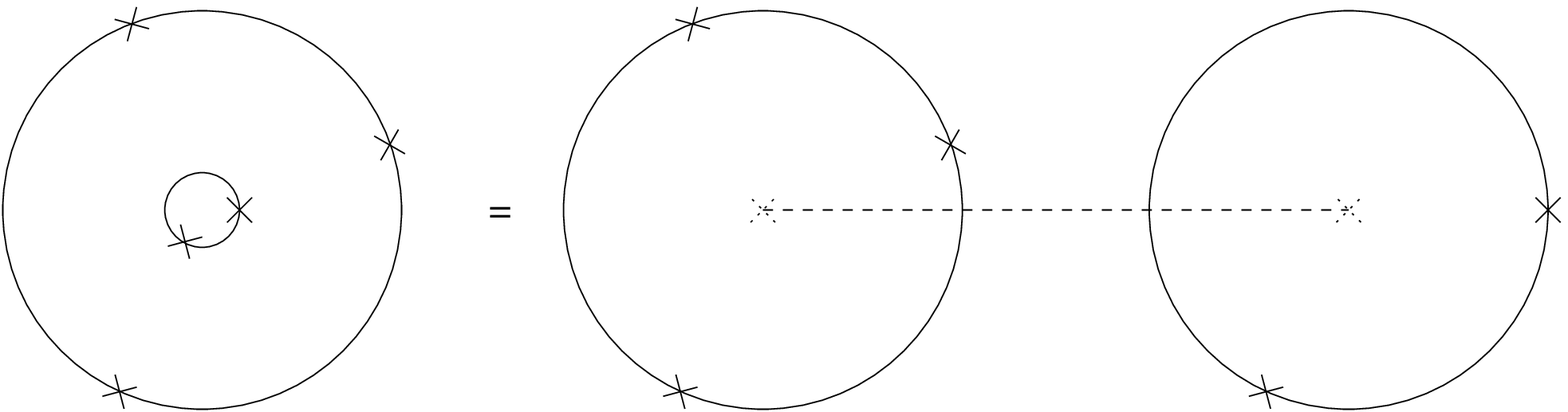}
\caption{The closed string factorization of the annulus into two
disks.\label{clfac}}
\end{center}
\end{figure}
\noindent where $S_{D_2}$ is the disk amplitude with an extra tachyon
insertion, and $K_{c}$ is the normalization of the tachyon two-point
function on the sphere.
Since
\begin{align} \label{roughd}
S_{\text{annulus}} &\sim  \sqrt{\det G}, &
S_{\text{disk}} &\sim {1 \ov g_o^2} \sqrt{\det G}, &
K_{c}  &\sim \frac{1}{g_c^2} \sqrt{\det g},
\end{align}
from~\eqref{ampcld2d2} and~\eqref{roughd} we find that
\begin{equation}
g_c = 2^{-18} \pi^{-25/2} {\alpha'}^{-6} \, g_o^2
   \left( \frac{\det g}{\det G} \right)^{\frac{1}{4}}
 = 2^{-18} \pi^{-25/2} {\alpha'}^{-6} \, g_o^2
   \left( \frac{\det g}{\det(g + B)} \right)^{\frac{1}{2}}.
\end{equation}
where the numerical factors are the same as in the $B=0$ case~\cite{jp}.

\section{IR/UV and Stretched Strings} \label{sec:iruv}

In the above we have shown that at the one-loop string theory level, the IR/UV
factor~\eqref{importf} is a pure open string effect. It has a very different
origin from the seemingly similar factor~\eqref{clofa} in the closed string 
channel, which arises only after summing over an infinite number of
open string
oscillation modes. As we emphasized above, equation~\eqref{importf} is due
solely
to the zero modes. Thus we conclude, at least in bosonic open string 
theory, that the IR/UV factor~\eqref{importf} is not related to the
appearance of
the closed string modes of the string theory in which the
noncommutative field
theory is embedded.   

Since we observed~\eqref{importf} occurring at the
perturbative 
string theory level in precisely the same manner as it does in noncommutative
field theories we might imagine that there is a simple explanation 
of~\eqref{importf} from string
theory. After all, constant $B$ is a rather mild background from
the string 
theory point of view, and we are supposed to know it very well. Indeed in
the following we will present a very simple explanation.  

 If we denote 
\begin{equation} \label{dx}
(\Delta x)^\mu = \Theta^{\mu \nu} p_\nu
\end{equation}
and interpret it as a distance,
then 
\begin{equation}
(\Delta x)^2 = - p_\mu (\Theta G \Theta)^{\mu \nu}  p_\nu = 
G_{\mu \nu} (\Delta x)^\mu (\Delta x)^\nu 
\end{equation}
has  the interpretation of a proper distance in terms of the
open string metric,
which indicates that there might be a stretched string interpretation
of~\eqref{importf}. This is supported by the fact that
in the background $B$ field, 
$X$ has the following mode expansion:
\begin{equation} \label{modeex}
X^{\mu} (\tau, \sigma) = x_0^{\mu} + 2 i \alpha' p^{\mu} \tau
+ {1 \ov \pi} \Theta^{\mu \nu} p_\nu \, \sigma 
+ (\alpha')^{1 \ov 2}(\text{oscillators}).
\end{equation}
Notice that with non-zero momentum there is an additional winding term 
proportional to
$\Theta^{\mu \nu} p_\nu$.
This has the consequence that, in addition to the standard oscillator piece 
the open string now has a finite constant  stretched length,
\begin{equation} \label{winding}
X^{\mu}(\tau, \sigma=\pi) - X^{\mu}(\tau, \sigma=0)
= \Theta^{\mu \nu} p_\nu  + (\alpha')^{1 \ov 2}({\rm oscillators}) \ .
\end{equation}

\FIGURE{
\begin{tabular}{rc}
(a) & \includegraphics[width=1in]{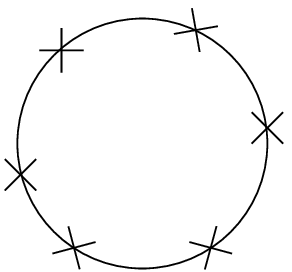} \\
(b) & \includegraphics[width=1in]{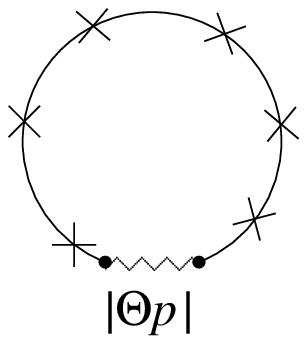}
\end{tabular}
\caption{First quantized picture of a one loop graph in field theory
for (a)~\hbox{$B=0$} and (b)~\hbox{$B\neq0$}.\label{1loop}}
}

Equation~\eqref{winding} is a beautiful manifestation of the UV/IR 
behavior of the system. When there is  non-zero momentum flow between 
the two boundaries of the string, as is the case for non-planar diagrams, 
there is an induced stretched string.
As an example, if we increase momentum in the $x^1$ direction by $\delta p_1$,
the string grows longer in the $x^2$ direction by an amount
$\abs{\Theta^{21} \delta p_1}$. As we increase the momentum further
the string stretches longer and longer. Since the first term in%
~\eqref{winding} has no $\alpha'$ dependence, 
the stretched length $\Theta^{\mu \nu} p_\nu$
persists in the field theory limit. It will act as an effective 
short-distance cutoff. As $p \rightarrow 0$, 
the stretched string length decreases 
to zero and we start seeing the standard short-distance  divergence again.

To understand the above picture more precisely, let us carefully 
go over the procedure
of taking the $\alpha'\rightarrow0$  limit. Figure \ref{cylinder} represents
a string world sheet with a coordinate width $\pi$ (and a space-time 
width in~\eqref{winding}) propagating through a proper time $2 \pi t$. 
When $B=0$, as we take the field
theory limit~\eqref{field}, the oscillation modes decouple and  
non-local effects 
caused by them in~\eqref{winding} also disappear and the worldsheet 
reduces to a circle of radius $\alpha' t$---see
figure~\ref{1loop}a. The circle is the
worldline  of a finite number of loop particles in the first quantized 
picture. The $\alpha' t \rightarrow 0$ limit amounts to shrinking the 
circle further to a point and is the usual short distance limit in field 
theory. In the second quantized 
picture $2 \pi \alpha' t$ may be identified as the Schwinger parameter
$\tau$---
\begin{equation} 
{1 \ov k^2 + m^2} = \int_{0}^\infty d\tau \, e^{- \tau (k^2 + m^2)}
\end{equation}
---and $\tau \rightarrow 0$ corresponds to the limit of high momentum. 
Now turn on the $B$-field and allow a non-zero momentum to flow 
through the worldsheet as for non-planar scattering. Due to the 
presence of the first term 
in~\eqref{winding}, the limit~\eqref{field} no longer shrinks
the world sheet to a circle, but instead to an arc
of width 
$\abs{\Theta^{\mu \nu} p_\nu}$---see figure~\ref{1loop}b. 
After taking  $\alpha' t \rightarrow 0$,
we are still left with a finite short-distance cutoff  of 
$\abs{\Theta^{\mu \nu} p_\nu}$.
Note that the above interpretation fits very well
with the way that~\eqref{importf} enters into the scattering 
amplitude~\eqref{ampop} ($\tau
= 2 \pi \alpha' t)$,
\begin{equation} \label{polya}
S  = \int d\tau \, \tau^{-\frac{d}{2} -1} \, e^{-{\Delta x^2 \ov 4  \tau}}
\times (\cdots)
\sim \int dt \, t^{-{d \ov 2} -1} \, \exp \big[
{1 \ov 8 \pi \alpha' t} p_\mu (\Theta G \Theta)^{\mu \nu}  p_\nu \big]
\times (\cdots)
\end{equation}
The structure in equation~\eqref{polya} is the usual structure for a
one-loop amplitude like figure~\ref{1loop}b in the first-quantized
approach to field theory; 
see {\em e.g.} ref.~\cite{ap}.

In the above discussion, the stretched strings appear to be rigid and 
non-dynamical; they merely cut open an otherwise closed particle loop. 
It is natural to ask whether this is  due to our looking only at 
the one-loop level, for, after all, equation~\eqref{modeex} is a
tree-level expansion.
The situation might change at higher genus; 
in particular, at the two-loop level, we will have vertices 
that are contracted only with internal lines. It is an interesting question 
whether the stretched string here is related to the ``noncommutative
QCD string'' discussed in~\cite{hi,mr,mrs}.   
In~\cite{mrs}
it was observed that the behaviour of higher genus diagrams suggests a
string coupling  of $g_s \sim 1/(E^2 \Theta)$ which 
seems to be consistent with supergravity solutions~\cite{hi,mr}.
Similarly, we observe that equation~\eqref{dx} is reminiscent of the
high energy proportionality between $x$ and $p$~\cite{gm} and thus
suggests $\alpha'_{\text{eff}} \sim \abs{\Theta}$ (see also~\cite{newiikk}).

\section{Discussion and Conclusions} \label{sec:conclu}

In this paper we have examined in detail the one-loop non-planar
string amplitude for $M$ tachyons  in bosonic open string theory
with a $B$-field. We showed that the IR/UV mixing behavior of~\cite{mrs} 
can be understood as a pure open string effect in terms of the
stretched strings.
The mechanism is rather simple and 
general; in particular it simultaneously explains the various
singularities (quadratic, logarithmic, etc.) of all operators
(two-point, four-point, etc.).

In our approach, the IR singularities observed in~\cite{mrs} are
not attributed to missing light degrees of freedom in the theory. They
are rather a disguised reflection of the UV behaviour of the theory.
The non-locality of the theory introduces a dynamical short-distance cutoff
which is proportional to the external non-planar momentum $p$.   
The $p \rightarrow 0$ limit simply removes this cutoff 
and recovers the UV divergences. 

This raises the question as to whether (bosonic)
noncommutative theories are consistent.%
\footnote{Here we mean whether the theory has a consistent 
{\em Wilsonian\/} description at all momenta.}
As already pointed out in~\cite{mrs}, if we impose a cutoff $\Lambda
=1/\epsilon$
into the  theory then for $(\Delta x \equiv \Theta p) < \epsilon$ the
limit of
taking $\Lambda\rightarrow\infty$ is not well-defined since there 
is a crossover between $\Delta x$ and $\epsilon$. 
For example, if we consider a noncommutative scalar $\Phi^3$ theory in six 
dimensions, or any other theory that is renormalizable for $\Theta=0$, 
embedded in a string theory, then the natural cutoff of the theory 
is $\epsilon=(\alpha')^{1/2}$. Let 
us first assume $\Theta > \alpha'$, where here and in the following,
by $\Theta$ we mean, roughly, the magnitude of a typical
eigenvalue. There are now
two ways of ``detecting'' the effects of the massive string modes. One 
is obvious; they are excited at high energies
of order $(\alpha')^{-1/2}$.
The other is to go to very low energies of 
order $(\alpha')^{1/2}/\Theta$.
More explicitly, inside the energy region
\begin{equation} \tag{region I}
(\alpha')^{1/2}/\Theta < p < (\alpha')^{-1/2}, 
\end{equation}
the one-loop
effective action has the heuristic form
\begin{equation} \label{highen}
\Gamma(k) = \Gamma_{\text{planar}}(k, (\alpha')^{1/2}) + 
 \Gamma_{\text{non-planar}} (k, \Theta p) 
\end{equation}
where $k$ denotes generic external momenta, $p$ non-planar momenta,
and the second variable denotes the cutoff. The non-planar amplitudes 
are finite; planar diagrams can be renormalized 
in the usual sense; and the $\alpha' \rightarrow 0$ limit can be consistently 
defined. However as we reach the region of small momenta,
\begin{equation} \tag{region II}
p<(\alpha')^{1/2}/\Theta,
\end{equation}
the effective action becomes
\begin{equation} \label{lowen}
\Gamma(k) = \Gamma_{\text{planar}}(k, (\alpha')^{1/2}) + 
 \Gamma_{\text{non-planar}}(k,  (\alpha')^{1/2}) 
\end{equation}
{\em i.e.} $\alpha'$ effects will take over below 
energy scales of order $(\alpha')^{1/2}/\Theta$.
This crossover can be seen explicitly in 
equation~\eqref{modeex}; below this scale 
the oscillator terms become important compared to the
$\Theta^{\mu\nu} p_\nu \sigma$ term.
Now, unlike for
ordinary field theories, it is no longer consistent to take the
$\alpha' \rightarrow 0$ limit since the behaviour of the full theory, when
$(\alpha')^{1/2}$ becomes smaller than $\Theta p$, is such that the
effective theory of equation~\eqref{lowen} is 
no longer valid, but rather we go over to the region I
behaviour~\eqref{highen}. We should caution that this discussion
is only at the one-loop level. At the two-loop level the loop integration 
over $p$ will  result in divergent diagrams, a reflection of the 
one-loop IR singularities.

Thus it appears that although na\"{\i}vely---as we have seen in
section~\ref{sec:green}---the
massive modes seem to decouple in the amplitude in the 
limit~\eqref{field}, in fact
the oscillators are quite important.
What is surprising is that they preclude a proper Wilsonian description
at {\em low\/} non-planar momentum! 
Only if we keep
$\Theta < \alpha'$ as $\alpha'  \rightarrow 0 $ do we recover
Wilsonian field theory, but it is then a commutative field theory.
Thus, for a noncommutative theory, we see here an intricate interplay between 
two scales $\Theta$ and $\alpha'$.

This is not to suggest that we cannot add new fields, 
along the lines of~\cite{mrs,rs},
to reinterprete the IR singularities.  Rather, we point
out that this interpretation of the singularities is not one
that arises naturally out of bosonic string theory.

\acknowledgments

We had useful conversations with O. Aharony, I. Brunner, M. Douglas,
K. Intriligator, G. Moore,
B. Pioline, M. Rozali, C. Zhou,
and especially A. Rajaraman and J. Shapiro. H.L. also wants to thank
A. Tseytlin for
emphasizing, very early on,
the importance of understanding the one-loop amplitude.
This work was supported by DOE grant
\hbox{\#DE-FG02-96ER40559} and an NSERC PDF fellowship.

\end{document}